\documentclass[aps, prd, twocolumn, lengthcheck, superscriptaddress, % showpacs,
nofootinbib]{revtex4-1}

\usepackage{epsfig}
\usepackage[usenames]{color}
\usepackage{graphicx}
\usepackage{amsmath}
\usepackage{epstopdf}

\def\bi{\bibitem}

\def\la{\langle}
\def\ra{\rangle}
\def\be{\begin{eqnarray}}\def\ee{\end{eqnarray}}
\def\lsim{\mathrel{\rlap{\lower3pt\hbox{\hskip1pt$\sim$}}
     \raise1pt\hbox{$<$}}} %less than or approx. symbol
\def\gsim{\mathrel{\rlap{\lower3pt\hbox{\hskip1pt$\sim$}}
     \raise1pt\hbox{$>$}}} %greater than or approx. symbol
\def\del{\partial}

\allowdisplaybreaks

\def\Tr{\rm Tr}
\def\bi{\bibitem}

\newcommand{\CL}{{\cal L}}

\def\B0{{\cal B}^{(0)}}
\def\K{{\phi}^\ast}

\begin{document}

%\title{Hidden Symmetries, Quantized Hall Droplets and Vector Meson Dominance \\ in Compressed Baryonic Matter}
\title{Skyrmions and Fractional Quantum Hall Droplets
\\ Unified  by Hidden Symmetries in Dense Matter}

%\author{Yong-Liang Ma}
%\email{ylma@ucas.ac.cn}
%\affiliation{School of Fundamental Physics and Mathematical Sciences,
%Hangzhou Institute for Advanced Study, UCAS, Hangzhou, 310024, China}
%\affiliation{International Center for Theoretical Physics Asia-Pacific (ICTP-AP) (Beijing/Hangzhou), UCAS, Beijing 100190, China}

\author{Mannque Rho}
\email{mannque.rho@ipht.fr}
\affiliation{Institut de Physique Th\'eorique, Universit\'e Paris-Saclay, CNRS, CEA,  91191, Gif-sur-Yvette, France }

\date{\today}

\begin{abstract}

A chain of connections in compressed baryonic matter, up-to-date missing in nuclear effective field theory, between emergent symmetries of QCD, mesons-gluons dualities, vector meson dominance and Chern-Simons fields  has recently been revealed, presaging a possible new paradigm in nuclear theory. It indicates an ubiquitous role, thus far unexplored,  of hidden flavor-local and scale symmetries  permeating from dilute baryonic systems to normal nuclear matter and then to compact-star matter. Here I give a brief, somewhat speculative, account of the possibly ``indispensable" relevance of the $\eta^\prime$ singular ring --  a.k.a. fractional quantum Hall (FQH) droplet -- to the properties of the lowest-lying vector mesons $\omega$ and $\rho$  considered to be Seiberg-dual to the gluons near the chiral restoration.  
\end{abstract}
\maketitle
\setcounter{footnote}{0}
\section{Introduction}
There has been a  dichotomy in the topological description of baryons in the large $N_c$ QCD that while the $N_f\geq 2$ baryons are skyrmions, the $N_f=1$ baryon cannot be a skyrmion. This dichotomy has recently been resolved by that the $N_f=1$ baryon is a fractional quantum Hall (FQH) droplet (or a ``pancake")~\cite{zohar}. This dichotomy could also be seen in the chiral bag model~\cite{chiral-bag} in the way the Cheshire Cat Principle works~\cite{MNRZ}: the skyrmions for $N_f\geq 2$ are interpreted in terms of the quarks dropping into an ``infinite hotel" in (3+1) dimensions~\cite{CCP,infinite-hotel} whereas the  FQH droplet for $N_f=1$ results from the anomaly-inflow mechanism with the quarks leaking from the bag into (2+1) dimensional droplet~\cite{anomaly-inflow}. The two apparently different pictures, it turns out, can be reconciled, or more appropriately ``unified," by hidden symmetries presumed to be intrinsic in QCD. In particular the hidden local symmetry (HLS) for the vector mesons $V=\rho,\omega$ familiar in hadronic interactions at low energy is seen to play the key role in the unification~\cite{karasik,karasik2,kitano,Y}. An observation which could have an important impact on nuclear physics under extreme conditions is that the properties of the vector mesons $V$ are correctly described only with the FQH droplets taken into account, hence both the HLS mesons and the FQH droplets play an indispensable role for the matter approaching the chiral phase transition at high temperature and/or density. 

In this paper I implement the spontaneously broken scale symmetry with a dilaton $\chi$ as a ``conformal compensator" (defined below) to the HLS Lagrangian. In contrast to what's done in Refs.\cite{karasik,karasik2,kitano} in terms of the dynamical domain walls (DWs) arising from mixed  't Hooft anomalies -- but without scale symmetry, my argument will hang crucially on the hidden scale symmetry.  This is because nuclear matter density  is exploited to drive the system.  Conformal symmetry or more specifically scale symmetry in gauge theory is still a highly controversial issue. So I will avoid the technical details. How the symmetry figures in the problem at hand will be addressed  in EFTs focused on its emergence at a variety of different length scales involved in nuclear processes, without asking how the manifestations at different scale are interconnected in the fundamental theory, QCD.

The strategy I adopt is to start from the nuclear effective field theory, dubbed  $Gn$EFT, defined concisely below and detailed in \cite{MR-review}  that generalizes the standard chiral EFT that is successful at nuclear matter density $n_0\sim 0.16$ fm$^{-3}$ to much higher densities relevant to  compact star properties and possibly beyond.  As described in Section \ref{remarks}, an emergent hidden scale symmetry is found to permeate from the unitarity limit at very low density to the nuclear matter density, then to the  neutron-star density and finally to the density at which the putative FQH configuration {\it inevitably} appears near the chiral restoration.

%%%%%

\section{Fractional Quantum Hall Droplets and Cheshire Cat}
%The key observation pertinent to the issue is that the baryon for the $N_f=1$ system cannot be a skyrmion  whereas all baryons for $N_f \geq  2$ (like nucleon, hyperons etc) do arise from octet mesons $\pi$ as skyrmions at large $N_c$ limit. 
It was suggested by Komargodski~\cite{zohar} that the one-flavored baryon (denoted  $\B0$) is a fractional quantum Hall (FQH) droplet associated with the flavor singlet meson $\eta^\prime$ in (2+1) dimensions, and not a skyrmion in (3+1) dimensions. It is not a skyrmion because $\pi_3 (U(1))=0$ contrary to $\pi_3 (SU(N_f)_{N_f \geq 2})={\cal Z}$. It is argued that the $\B0$ is a domain-wall configuration of the flavor singlet meson $\eta^\prime$ bounded by a string in the form of a pancake (or pita) on which $U(1)_{-N_c}$ Chern-Simons theory is supported with the edge modes encoding the baryon number and the spin,  $J=N_c/2$.
%\footnote{Note that $\eta^\prime$ in nature is an octet-singlet mixture of flavor $U(3)$. Here we are dealing with the singlet $\in U_A(1)$.}  
 
For what follows, I find it more appropriate to follow the Cheshire Cat Principle (CCP) using the chiral bag model in describing the difference between the $\B0$ and the nucleon $N$~\cite{MNRZ}. 

In the chiral bag model~\cite{chiral-bag}, the nucleon is described by $N_c=3$ nearly massless quarks in color singlet confined inside an MIT bag of radius $R$ clouded outside by the Nambu-Goldstone bosons $\pi$. The quarks in the bag and the pion fields outside the bag are connected by the conserved vector and axial-vector currents. The boundary conditions induce a quantum anomaly in the quark vector current, so the baryon charge leaks out of the  bag. The anomaly cancellation takes place with the leaked baryon charge picked up by the soliton outside.  This phenomenon can be described in (1+1) dimensions in terms  of the baryon charge dropping into an ``infinite hotel (IH)" having  to do with regularizing infinity~\cite{infinite-hotel}. This can be worked out analytically in (1+1) dimensions.   When the bag radius $R$ is shrunk to zero, the baryon charge gets lodged entirely in the cloud supporting a skyrmion. This simple picture holds realistically in (3+1) dimensions~\cite{goldstone-jaffe}.  The deep import of this result is that the bag radius, interpreted in the MIT bag model as the ``confinement radius," is in fact unphysical, much like a gauge parameter of gauge theories.  This notion  that the confinement size is a gauge parameter~\cite{gauge} is the Cheshire Cat Principle (CCP for short)~\cite{CCP}. 

\section{Dichotomy Problem}
Suppose now we have $N_c$ one-flavor quarks ``confined" in the bag  coupled chiral-symmetrically to the $\eta^\prime$ mesons living outside. Consider  what happens to the  quarks propagating in (1+1) dimensions. The (chiral) boundary conditions break the current conservations involving $\eta^\prime$ as in the case of $N_f=2$, so the baryon charge must also leak out.  How does this leakage take place? The $\infty$-hotel mechanism does not work in this case because  $\pi_3(U(1))= 0$. Instead the anomaly so generated is compensated by the quarks flowing in one dimension higher, i.e.,  in 2 space dimensions, thereby forming a sheet. It supports abelian Chern-Simons theory. This is consistent with the well-known  ``anomaly inflow" mechanism in (2+1) dimensions~\cite{anomaly-inflow}. By construction, the resulting Chern-Simons droplet depends on the size $R$, hence non-gauge-invariant for an arbitrary $R<\infty$.  The non-gauge-invariance must be compensated by the quarks. When the baryon charge is completely leaked into the droplet, the Cheshire-Cat ``smile" resides in a vortex line. This  precisely reproduces the CCP.  

There is a caveat here. Naively applied to, say,  $N_f=2$, there is no known reason why the two-flavored quarks could not make the anomaly-inflow to a non-abelian Chern-Simons action~\cite{MNRZ}. The question then is what determines the $N_f=2$ quarks to drop in the $\infty$-hotel, skyrmions, or flow to nonabelian QH droplets?
These two ways of baryon charge leakage present one aspect of the  ``dichotomy problem (DP)."  There are several other issues associated with this DP as discussed in \cite{DP}.  At present, one can think of a few possibilities. I will discuss however only one particular aspect that could be relevant to high-density or high-T matter.

The issue I will address is: Is the baryon charge of the $\B0$  connected to that of the $N$ and if so, how?  Off-hand the baryon charges of the two objects look totally disconnected. But  surprisingly, the two seem in fact to be tightly interwoven.   In particular, the famous but mysterious notion of ``VMD (vector meson dominance)"  which has played an important role in hadronic electroweak form factors, more specially in nuclear physics,  is found to link the $\B0$ and $N$.  Most important in providing the link  is the role played by {\it hidden local symmetry (HLS)} together with {\it hidden scale symmetry}  (implemented in nuclear medium)  to provide a simple unified framework to describe baryonic matter from normal nuclear matter to compact star matter~\cite{MR-review}.

\section{Hidden Local Symmetry,  Chern-Simons Fields and Vector Meson Dominance}
As first proposed in \cite{karasik,karasik2} and then in \cite{kitano}, the FQH droplets and skyrmions could be unified  by identifying the hidden local vector fields with the Chern-Simons fields via a Seiberg-type duality.  The key point there is that when $\eta^\prime$ is coupled into HLS Lagrangian, the $U(1)$, i.e., $\omega$, can be identified with the Chern-Simons abelian gauge field. The $\rho$ fields can be added to extend to the $N_f=2$ case, i.e., nonabelian CS theory.\footnote{There seems to be certain differences between \cite{karasik,karasik2} and \cite{kitano} in the way the Chern-Simons field(s) is(are) identified with HLS field(s). My discussion will bypass the possible differences by going over the intermediate steps involving different density regimes in dense matter. I will roughly follow the line of reasoning of Karasik.}  
To start with, put the $\eta^\prime$ meson together with the pions in the chiral field $U\equiv \xi^2$, 
\be
U=e^{i\eta^\prime} e^{i\tau_a\pi_a/f_\pi}.
\ee
For simplicity,  consider the 2 flavor (u, d) system\footnote{Generalization to the 3-flavor system is straightforward.}. 
It was shown in \cite{karasik} how by  ``tweaking"  quark masses, one can  distort the hedgehog configurations of the  skyrmions in $U$  to a configuration in which $\eta^\prime$ winds around a singular ring of the FQH droplet $\B0$.  For this,  ``dial" the d-quark mass $m_d$ to $\infty$ while keeping $m_u=0$~\cite{karasik}. A new heavy degree of freedom is seen to emerge  on the ($\eta^\prime$) singular ring and  gives rise to the FQH droplet with the heavy degree of freedom appearing as chiral edge mode. This heavy degree of freedom is then identified with the Chern-Simons field. This information is encoded in what's generically called  ``hWZ" term in HLS theory. It  is the parity-anomalous Wess-Zumino Lagrangian that contains the vector mesons~\cite{HLS84,HY:PR} that satisfies, together with the 5-dimensional term, the Wess-Zumino anomaly equation.   It is referred to as ``homogeneous" Wess-Zumino term in \cite{HLS84,HY:PR} and as ``hidden" Wess-Zumino term in \cite{karasik,karasik2}. The difference between the two is essential for what follows.
How the connection between the FQH configuration and the skyrmion configuration is encoded in this hWZ term is explicitly shown in \cite{karasik,karasik2}. 

Now what can one tweak in nuclear physics to go from one to the two configurations?  Tweaking the quark mass is obviously not feasible. I will tweak the baryon density as one does in the EoS for compact-star physics. It is here that the hidden scale symmetry figures importantly in the EoS.

What is noteworthy about the HLS that is adopted in the $Gn$EFT approach~\cite{MR-review}  is that the vector fields $V_\mu=(\rho_\mu,\omega_\mu)$ are taken as ``dynamically generated" gauge fields~\cite{bando-kugo-yamawaki,suzuki}. This implies that in the chiral limit, there is  ``vector manifestation (VM)" hidden in the sense that at some point in high density, say, $n_{\rm VM}\gsim 25 n_0$,  or at high T, say, at $T_c\approx 150$ MeV where chiral symmetry is presumably restored, the gauge coupling $g_V$ goes zero such that $m_V^2\to 0$~\cite{HY:PR}. This is highly relevant to the behavior of the vector mesons near chiral restoration as in dilepton production experiments in heavy-ion collisions. I will return to this matter below.

In Karasik's analyses, scale symmetry does not figure. That's because it is subdominant in the large $N_c$ counting in the (matter-free) vacuum. It does come in at higher tracing, but  suppressed by $1/N_c$. We will see, however, that the effect of scale symmetry is unsuppressed in dense baryonic matter. 
\subsection{HLS at leading order}
Before going into dense matter, consider what HLS encodes.

What underlies the reasoning presented  is the possible Seiberg duality of HLS to the QCD gluons. At present it's only a conjecture based on analogy to ${\cal N}=1$ supersymmetric YM theory, not yet given a proof, but there is a growing evidence that the duality conjecture is  valid in strong interactions~\cite{Komargodski,abel-bernard,Y}. What's remarkable is that HLS at the leading order in the power expansion, i.e., $O(p^2)$, has a magical power with certain characteristic property of SUSY~\cite{Komargodski}. Now what's found by Karasik in connection with the FQH droplet structure is that HLS plays the role of ``heavy degrees of freedom" involved with the $\eta^\prime$ singularity, so when the HLS vectors are {\it integrated in} the chiral Lagrangian for the $\eta^\prime$, the singularity is eliminated in the potential.

Let's first write  the relevant HLS Lagrangian  to $O(p^2)$. Define the HLS fields as
\be
V_\mu^\rho& = &{} \frac{1}{2}g_\rho \rho_\mu^a\tau^a, \quad V_\mu^\omega = {} \frac{1}{2}g_\omega \omega_\mu.
\ee
and  the Maurer-Cartan 1-forms
\be
\hat{\alpha}_{\parallel,\perp}^\mu & = & \frac{1}{2i}\left(D^\mu \xi \cdot \xi^\dagger \pm D^\mu \xi^\dagger\cdot \xi\right)
\ee
where $D_\mu \xi = (\partial_\mu - iV_\mu^\rho - i V_\mu^\omega)\xi$.

The HLS is given in the form
\be
{\CL}_{HLS}
& = & f^2\text{Tr}\left(\hat{\alpha}^\mu_\perp\hat{\alpha}_{\perp\mu}\right) + f_{\sigma\rho}^2\text{Tr}\left(\hat{\alpha}^\mu_\parallel\hat{\alpha}_{\parallel\mu}\right)\nonumber\\
& &{} + f_{0}^2\text{Tr}\left(\hat{\alpha}^\mu_\parallel\right)\text{Tr}\left(\hat{\alpha}_{\parallel\mu}\right) \ + {\CL}_{hWZ\nonumber}\\
& &{} - \frac{1}{2g_\rho^2}\text{Tr}(V_{\mu\nu}V^{\mu\nu}) - \frac{1}{2g_0^2}\text{Tr}(V_{\mu\nu})\text{Tr}(V^{\mu\nu}) 
\label{hls}
\ee
%\be
%V^{\mu\nu} & = & \partial^\mu V^\nu - \partial^\nu V^\mu - i [V^\mu, V^\nu]
%\ee
with $V^\mu = V_\rho^\mu + V_\omega^\mu$. The presence of the various parameters of the Lagrangian such as $f, f_0, f_{\sigma\rho}, g_{0,\rho}$ is due to the fact  the $\rho$ and $\omega$ are treated not in $U(2)$ but in $SU(2)\times U(1)$.

The hWZ Lagrangian that will be found to unify the $N_f=1$  and $N_f=2$ baryons\footnote{From here on, I will consider $N_f=2$ instead of $N_f\geq 2$.} consists of four terms coupled to the external $U(1)_{\rm ext}$ field, either baryonic $B_\mu$ or EM $J_\mu^{\rm em}$,
\be
{\CL}_{hWZ}= \frac{N_c}{16\pi^2}\sum _{i=1}^{4} c_i {\CL}_i.\label{hWZterm}
\ee
The ${\CL}_i$s are constructed gauge-invariantly in both HLS and $U(1)_{\rm ext}$ with  the Maurer-Cartan 1-forms. The fourth term ${\CL}_4$ is absent in the absence of $U(1)_{\rm ext}$.    I skip writing the explicit forms of the terms.   They are neither transparent nor essential for discussions that follow.  If needed, the details can be found in \cite{MR-review}.
\subsection{``Homogeneous" WZ terms from VMD}

Three out of the four coefficients in (\ref{hWZterm}) can be fit by demanding the processes   $\omega\to\pi\pi\pi$, $\pi^0\gamma\gamma$ and $\omega\to \pi^0\gamma$ to be entirely given by the vector meson dominance (VMD). This leads to the prediction
\be
c_1-c_2=c_3=c_4=1.\label{VD-constants}
\ee
They are found to agree well with the experiments~\cite{HY:PR}. This then validates the VMD in the dynamically generated homogeneous WZ term. This is entirely consistent with  the VMD for the mass formula $m_V^2=a f_\pi^2 g_V^2$ with $a=2$ for the $\gamma\pi$ coupling in HLS and with what one deduces from the conjectured Seiberg duality~\cite{Komargodski,abel-bernard}. Note that one coefficient $c_1$ (equivalently $c_2$) remains unfixed by the VMD.

\subsection{``Hidden" WZ terms from FQH droplet}

Let's turn to what the FQH droplet for $N_f=1$ requires for $c_i$s in the hidden WZ.

Looking at the terms in(\ref{hWZterm}) that involve couplings to $\eta^\prime$, one gets the baryon current
\be
B^{(N_f=1)}=-\frac{c_3+c_4}{8\pi^2} d\omega d\eta^\prime.
\ee
This shows first that the  $\omega$ field can be identified with the abelian Chern-Simons field. Furthermore
Since 
\be
-\frac{1}{4\pi^2}\int_{M^3} d\omega d\eta \in Z
\ee
it follows that $B=1$ leads to 
\be
c_3+c_4=2\label{c3-c4}.
\ee 
This,  what could be called as a quantization condition,  reproduces  the VDM value (\ref{VD-constants}).

Now applying the rank-level duality $SU(N_c)_{1}\simeq U(1)_{- N_c}$, identifying the $\omega$ with the Chern-Simons field on the $\eta^\prime$ domain wall (DW) leads to 
\be
c_3=c_4=1\label{c3}
\ee
which confirms (\ref{c3-c4}).

Finally to determine the remaining coefficient $c_1$ or $c_2$, one has to look at $N_f=2$ baryons in the presence of $\eta^\prime$ because ${\CL}_1 ={\CL}_2 =0$ for $N_f=1$. 
The calculation is a bit involved. It turns out~\cite{karasik2,Y} that it involves nonabelian Chern-Simons fields -- since $N_f=2$ --  given in 3-form by
\be
{\cal L}_{CS\eta^\prime}= - \frac{N_c}{8\pi^2} d\eta{\Tr} \big(V dV +\frac{2}{3}V^3\big)\label{etaCS}
\ee
where $V_\mu=\frac 12 (\tau\cdot\rho_\mu+\omega_\mu)$. One gets
\be
c_1=2/3.
\ee 
\subsection{Color anomaly and Cheshire Cat Priniciple}
It is interesting to note that  if one applies the rank-level duality $SU(N_c)_{N_f}\leftrightarrow U(N_f)_{-N_c}$ to (\ref{etaCS}),  this Lagrangian, expressed in terms of gluon fields for $V_\mu$ is  precisely the same as the counter term that figures  on the surface of the chiral bag model to cancel the color leaking out of the bag~\cite{coloranomaly}.  A  support for this identification~\cite{FSAC} comes from the famous ``proton-spin problem," namely the  observed  flavor-singlet axial constant (FSAC) $g_A^{(0)} <<1$ in stark contrast to what's predicted  $g_A^{(0)}\sim 1$ by the constituent quark model.  This suggests the chiral bag model should include HLS fields as edge modes argued in \cite{kitano} in addition to the pions excited outside of the bag. This would require a new formulation of the chiral bag model

To summarize,  the VDM ``unifies" the FQH droplet baryons and the skyrmion baryons.  There will be  more to say on this matter at the end of this note.

\section{Hidden Scale Symmetry and genuine dilaton}
Up to this point considered in terms of domain walls, scale symmetry played no role. However in the presence of baryonic matter, scale symmetry becomes indispensable for dialing the system. To see how scale symmetry can figure, let's consider the $Gn$EFT that has been found to fairly successfully describe baryonic matter up to compact star density $\sim (6-7)n_0$ reviewed in \cite{MR-review} and see whether one can approach the FQH droplet configuration at a density where chiral symmetry could be restored. In the theory described in \cite{MR-review}, the dilaton decay constant remains nonzero in massive neutron stars, hence chiral-scale symmetry is spontaneously broken, with the energy-momentum tensor not traceless and yet gives the sound speed pseudo-conformal $v_{pcs}^2/c^2\approx 1/3$ in neutron stars of mass $M\approx 2 M_\odot$. In this system, the FQH droplet configuration is either merely invisible or simply unstable. The question is: By going up further in density, can one ``expose" the FQH droplet configuration? 
%
%\subsection{``Genuine dilaton"}
%
Following the approach that borrows from ${\cal N}=1$ SUSY  to incorporate the putative mesons-gluons duality,  scale symmetry is introduced with what's called ``genuine dilaton" (GD for short) for the scalar labeled $f_0(500)$ in the particle booklet as suggested by Crewther~\cite{crewther,CT}. In this GD, it is posited that there is an infrared fixed point (IRFP) $\alpha_{\rm IRs}$ at which the $\beta$ function vanishes (in the chiral limit),
\be 
 \beta_{QCD} (\alpha_s = \alpha_{\rm IRs})=  0 
\ee
where $\alpha_s$ is the QCD coupling constant $\frac{g_s^2}{4\pi}$.  At the IRFP, the dilaton joins the Nambu-Goldstone bosons $\pi, K, \eta$, with non-vanishing dilaton condensate, acting as a scale condensate.  The dilaton mass goes as $m_\sigma^2 \propto (\alpha_{\rm IRs}-\alpha_s)\beta_{QCD}^\prime$  with $\beta^\prime >0$ in the chiral limit, so the non-zero dilaton mass in the real world, e.g., neutron stars, indicates the departure of $\alpha_s$ from the IRFP value~\cite{crewther}
\be
\alpha_{\rm IRs}-\alpha_s \simeq \alpha_s (0)-\alpha_s(-m_K^2)=O(m^2_K/m^2_\pi).
\ee
Spontaneous scale symmetry breaking endows -- with non-zero condensates -- masses to matter fields, i.e., nucleons, vector mesons etc. In the treatment at hand, one cannot -- need not -- enter into the basic structure of the GD notion but the idea, in a nut-shell, is similar to the notion of ``quantum scale invariance" that addresses, among others, the cosmological constant~\cite{QSI} where the mass scale $\mu$ that intervenes in renormalization, breaking scale invariance explicitly,  is lifted to a scale-dimensional field operator, namely, the ``conformal compensator (CC)."  This procedure, which is consistent with the exploitation of the mesons-gluons duality~\cite{abel-bernard},  trades the standard renormalization of renormalizable theories (such as QCD) for scale-invariance at the quantum level bypassing the issue of standard renormalizability. The upshot is that in this way scale invariance is {\it hidden} at the tree level but {\it emerges} at the quantum level. There is nothing wrong with this procedure given that one is dealing here with an EFT. This accounts for the term ``hidden scale symmetry." In many-body nuclear processes, this applies both to the fundamental (i.e., QCD) and nuclear renormalizations. 

The prescription to construct the GD-scale-symmetry implemented HLS (dubbed as $\chi$HLS\footnote{Here and in what follows, $\chi$ stands for the conformal compensator field employed for the genuine dilaton, not to be confused with chiral symmetry.} ) is simple. Multiply the appropriate power of the CC field $\chi=f_\chi e^{\sigma/f_\chi}$  (that transforms linearly under scale transformation whereas $\sigma$  transforms non-linearly)  so as to render  $\chi$HLS scale-invariant. The resulting $\chi$HLS Lagrangian takes the form
\be
{\CL}_{sHLS}
& = & f^2\Phi^2\text{Tr}\left(\hat{\alpha}^\mu_\perp\hat{\alpha}_{\perp\mu}\right) + f_{\sigma\rho}^2\Phi^2\text{Tr}\left(\hat{\alpha}^\mu_\parallel\hat{\alpha}_{\parallel\mu}\right)\nonumber\\
& &{} + f_{0}^2\Phi^2\text{Tr}\left(\hat{\alpha}^\mu_\parallel\right)\text{Tr}\left(\hat{\alpha}_{\parallel\mu}\right) \ + {\CL}_{hWZ\nonumber}\\
& &{} - \frac{1}{2g_\rho^2}\text{Tr}(V_{\mu\nu}V^{\mu\nu}) - \frac{1}{2g_0^2}\text{Tr}(V_{\mu\nu})\text{Tr}(V^{\mu\nu}) \nonumber \\
& &{} +\frac{1}{2}\partial_\mu \chi \partial^\mu \chi + V_d(\chi)
\label{shls}
\ee
where $\Phi$ is the conformal compensator field 
\be
\Phi\equiv \chi/\la\chi\ra_0
\ee
where $\la\chi\ra_0$ is the dilaton condensate $\propto f_\chi$ with $f_\chi$ the dilaton decay constant in the matter-free vacuum, and $V_d(\chi)$ is the dilaton potential  that triggers scale-symmetry spontaneous breaking. Note that the vector kinetic terms and the hWZ term, both scale-invariant at the classical level, are unaffected by $\Phi$.  It is important for what follows that both HLS and scale symmetry are implemented in (\ref{shls}) so that quantum renormalizations are taken into account at the tree level so that the mean-field approximation with $\chi$HLS  at the leading order chiral expansion as discussed in \cite{Komargodski,abel-bernard,Y} and the notion of quantum scale invariance as given in \cite{QSI} capture quantum effects. A specific example in nuclear physics is the famous long-standing quenching of the Gamow-Teller coupling constant $g_A$ from the free space value $g_A=1.276$ to the renormalized value $g_A^{\rm eff}\approx 1$ that illustrates how scale symmetry hidden at the tree level in nuclear interactions emerges after full nuclear correlations are taken into account~\cite{gA}.  This matter will be taken up below for the pervasiveness of hidden scale symmetry in baryonic matter from low to high density.

\section{Going Onto the $\eta^\prime$ Singular Ring}
I will now examine what happens to $\chi$HLS as density is increased. Embedded in the vacuum modified by the density, the parameters of the $\chi$HLS Lagrangian get endowed by the intrinsic density dependence (IDD) inherited from QCD at the scale at which the EFT and QCD are matched. This is an old story in nuclear physics dating  way back to 1991~\cite{br91}. The IDD that controls the properties of hadrons as density (or temperature) goes toward the critical value for chiral restoration plays a crucial role not only in normal nuclear matter but also in dense neutron-star matter~\cite{MR-review}. What turns out to be the most important for what follows is how the dilaton condensate in dense medium comes in to make the $\eta^\prime$ singular ring become visible as the matter approaches chiral restoration. The vacuum change requires that the CC field be shifted $\chi\to \la\chi\ra^\ast +\chi^\prime$ with the $\ast$ indicating density dependence. Dropping the fluctuating field $\chi^\prime$ which plays no leading role in what follows\footnote{As I will mention in Sect. \ref{remarks}, going beyond what corresponds to the Fermi-liquid fixed point approximation, one should be able to formulate a systematic scale-chiral perturbation approach in $Gn$EFT.}, the resulting $\chi$HLS Lagrangian takes the same form as (\ref{shls}) with $\Phi$ replaced by 
\be
\phi^\ast\equiv f_\chi^\ast/f_\chi. \label{Phistar}
\ee
In the GD scheme, the pion decay constant  is related to the dilaton decay constant, so not too far from the matter-free vacuum  (say near nuclear matter), $\phi^\ast$ can be taken as $f_\pi^\ast/f_\pi$, enabling the Lagrangian quantitatively checked against nuclear experiments. 
\subsection{Topology change and parity-doubling}
Going beyond the normal nuclear matter density is quite subtle due to the possibility that a sort of hadron-quark continuity must be  taking  place near $n_{\rm hqc}\sim (3-4) n_0$. This possibility is accounted for by the topology change traded-in for hadron-quark continuity~\cite{MR-review}. The details of how to go about doing this are not needed, so I will  sketch only the most essential steps needed to pass  beyond $n_{\rm hqc}$ and go near the presumed FQH droplet configuration.
 
 The topology change  is essential for compact-star physics because it provides the indispensable mechanism for the EoS to change from soft (nuclear matter) to hard (massive stars) to account for massive neutron stars. What enters here is the vector mesons in HLS to generate the tensor forces and repulsive core among nucleons and produce a baryonic state that roughly corresponds to a ``pseudo-gap phase" with the vanishing quark condensate $\la\bar{q}q\ra\sim \la\chi\ra=0$ but non-zero gap $f_\pi\sim f_\chi \neq 0$. This structure persists beyond the density relevant to the core of stars and account for the pseudo-conformal sound speed $v_{\rm pcs}^2/c^2\approx 1/3$ precociously setting in at non-asymptotic density $\sim 3 n_0$ as predicted in \cite{MR-review}.
 
 Next there emerges  the parity-douplet symmetry of the nucleons in conjunction with the pseudo-gap structure at $\sim 3 n_0$.  As $\la\bar{q}q\ra\to 0$, the nucleon mass tends to a chiral-scalar mass $m_0$ which is $\sim (0.6-0.9)m_N$ over the range of compact-star density $\sim (6-7) n_0$. The nucleon mass evolves as density increases
 \be
 m_N\propto f_\chi\to m_0.
 \ee
 It is not known what will happen to $m_0$ as the chiral restoration sets in, but it seems reasonable to expect it to vanish at the deconfinement transition.  
 \subsection{Dilaton-limi fixed point (DLFP)}
 Another observation, no less important, is the behavior of the Lagrangian $\chi$HLS (\ref{shls}) coupled to the nucleons.  For this, the baryon field $\psi$ is coupled to HLS and the Lagrangian is made quantum scale-invariant in the same way as done with $\chi$HLS. Call it $\psi\chi$HLS. I will return to it below.

 Now take this $\psi\chi$HLS Lagrangian and do  the field reparametrization $\Sigma\equiv U\chi f_\pi/f_\chi=s+i\tau\cdot\pi$ and let ${\Tr}\Sigma\Sigma^\dagger\to 0$. There emerge  two singular terms in that limit. Requiring those singular terms to be absent lead to what's called  DLFP  ``constraints"~\cite{bira}
 \be
 g_{\rho NN}\equiv g_\rho (g_{v\rho} - 1)\to 0
 \ee
 where $g_{v\rho}$ is the nuclear renormalization for the $\rho$-coupling to nucleons and
\be
 g_A\to g_{v\rho}\to 1,\ f_\chi\to f_\pi \neq 0.
 \ee
This implies that the $\rho$ decouples from nucleons before the $\rho$ mass goes to zero (via the vector manifestation~\cite{HY:PR} with $g_\rho\to 0$).  I point out here that the DLPF limit $g_A\to 1$ is the continuity at high density of the ``quenched $g_A$" observed in nuclei discussed in Sect. \ref{remarks}.

With the above effects taken into account, the Lagrangian  (\ref{shls}) reduces in going toward the putative chiral restoration density $c_{\rm CR}\sim n_{\rm VM}$ to
\be
{\CL}_{\chi HLS} &=& \frac{1}{2}\partial_\mu \chi \partial^\mu \chi + V_d(\chi)+\frac 12 {\K}^2 (\del_\mu \chi^\prime)^2\nonumber\\
&-& (\omega_{\mu\nu})^2 +\frac 12 {\K}^2 m_\omega^2  (\omega_\mu)^2\nonumber\\
&-& c_3 \frac{N_c}{8\pi^2} \epsilon^{\mu\nu\alpha\beta} \omega_\mu\del_\nu\omega_\alpha \del_\beta\eta^\prime+\cdots
\ee

Now to go to near the $\eta^\prime$ singular ring going beyond the DLFP, I jump over what happens in between and take the $\K\to 0$. The $\eta^\prime$ kinetic energy term which is ill-defined near the singular ring.  Likewise both the dilaton and the $\omega$, joining the $\rho$,  become massless. Now the last term taken on the $\eta^\prime$ singularity ring corresponds to a Chern-Simons theory with a boundary, which requires the presence of a chiral boson as an edge mode to assure gauge invariance of the theory~\cite{karasik}. Thus in this approach $f_\chi^\ast\to 0$ is demanded so as to reproduce the results of the FQH pancake of \cite{zohar}.  This fixes $c_3=1$ agreeing the VDM result (\ref{c3}). 

  Thus the $\eta^\prime$ singular ring emerges from (\ref{shls}) as the homogeneous  WZ term goes over to the hidden WZ term with the exact vector meson dominance. The rest of the arguments based on the DW consideration follow from $\chi$HLS. 
 \section{Fate of the HLS vector mesons}
 The $Gn$EFT approach, as formulated, cannot address what happens between the density regime relevant to compact-star physics and the density at which the $\eta^\prime$ singularity ring becomes visible.  Karasik's approach has shown that starting from the $N_f=2$ sector, the  skyrmion configuration,  $U=\sigma+i\tau_i\pi_i$ with $\sigma^2+\pi_i^2=1$, can be distorted to $\propto (\sigma+i\pi_3)/\sqrt{\sigma^2+\pi_3^2}$ and then flow to $\eta^\prime$ ring in $N_f=1$. 
The $Gn$EFT approach does however show what happens in the density regime of $\sim 2 M_\odot$ stars~\cite{MR-review}. There  $f_\chi^\ast\sim f_\pi^\ast\neq 0$, $m_\sigma\sim \la\theta_\mu^\mu\ra \neq 0$, so it is closer to the dilaton-limit fixed-point regime than to the GD's IRFP but far from the density at which the FQH pancake configuration is reached.   $\la\theta_\mu^\mu\ra \neq 0$, hence not conformal, but $\frac{\del}{\del n}\la\theta_\mu^\mu\ra = 0$, therefore the sound speed is pseudo-conformal $v_{\rm pcs}^2/c^2\approx 1/3$~\cite{MR-review}. While this has been confirmed neither by QCD proper nor by observations,  as far as I am aware, this prediction is up to date -- not at odds with nature and is certainly not ruled out, although there are at present no model calculations that seem to support it either.

What's the most interesting is the question of what happens to the HLS vector mesons, in particular the $\rho$ meson, as the chiral restoration is approached as $\K\to 0$. This issue is most likely irrelevant to stable neutron stars (the density in stars is much too low), but it will be highly relevant to relativistic heavy-ion collisions looking for dropping $\rho$ mass at high temperature. The search for the signal of dropping $\rho$ mass at near the critical temperature for chiral restoration has failed to see the signal and that led to the conclusion that the VM in HLS is ``ruled out"~\cite{dilepton}.  It was however pointed out by those working along the line consistent with the hidden local symmetry as described above -- and  it seems mostly unrecognized in the RHIC physics community -- that this conclusion was much too premature and in fact totally unfounded when analyzed in terms of renormalization-group flow arguments~\cite{needle}.  The signal searched in the experiments was likened to a ``needle in a giant haystack" that the experiments simply failed to see even if it existed there.   The role of the FQH droplets near chiral restoration~\cite{karasik2,kitano}, if confirmed to be valid, would forcibly support the ``needle-in-haystack" scenario. At high temperature, the dimensional reduction to 3d allows the mesons-gluons Seiberg duality to apply and near the chiral restoration there must take place a phase transition from the Chern-Simons-Higgs phase -- massive vector mesons -- to Chern-Simons topological phase~\cite{Y}.  The hidden local symmetry vector mesons in the topological phase are identified with the Chern-Simons fields dual to the gluons\footnote{There may be some difference in the arguments between \cite{karasik2} and \cite{Y} before entering into the topological phase. The latter invokes the background field in the bulk of the QH pancake and HLS fields at the edge of the pancake with the mixing between them controlled by gauge invariance. But both seem to agree what happens at the end of the day with which my argument is concerned.}.  The most crucial point is that the FQH pancakes turn out to be absolutely essential for how the vector mesons, which are now identified as Chern-Simons fields, behave at the phase transition. In \cite{needle}, the vector mesons were treated with the vector manifestation \`a la \cite{HY:PR} with modified photon-coupling, deviating drastically from the VMD. If the VMD persists up to the exposure of the $\eta^\prime$ as discussed, the scenario will be different from what's given in \cite{HY:PR}.   In short, the theoretical analyses that have entered in the dilepton experiments analyzed so far could have missed a highly non-negligible behavior of the vector mesons in the vicinity of the chiral phase transition.\footnote{There is a recent analysis in the QCD-sum rule technique approach to the vector-meson  mass in the chirally-restored (Wigner-Weyl) phase which indicates that the mass does not go to zero~\cite{SHL}. A similar observation is made by \cite{kitano} in their approach to the Chern-Simons structure of mesons on the domain wall. This issue needs ultimately to be clarified.}

This development raises interesting questions.

In  holographic QCD, the vector dominance is achieved only when the high towers of the vector mesons, in addition to what corresponds to the CS fields,  are taken into account. This is more so in the baryonic form factors~\cite{nucleonFF}: It is fairly well-established that the VMD does not work well in nuclear EM form factors, even at low momentum transfers. Furthermore while the hWZ terms are scale-invariant at the classical level, they can be affected by nonperturbative nuclear renormalizations at the quantum level. This has been seen in how the hWZ terms should be modified at high density so as to suppress the $\omega$ meson mass diverging as density increases~\cite{PRV,MR-omega}.  What seems to matter here is the role of scale invariance that differs between the DW (domain wall) consideration~\cite{karasik,karasik2,kitano,Y} and the CCP consideration.

In certain descriptions of the EoS in the core of massive neutron stars, the color-flavor-locked (CFL) phase is invoked~\cite{CFL}.  The CFL structure can be properly justified by perturbative QCD only at asymptotic density. In such an approach, even if the CFL phase becomes relevant, the structure with the $\eta^\prime$ ring could precede the CFL phase. It would be interesting to explore how the pancake structure impacts the CFL structure in the high density EoS.

\section{Permeating hidden scale symmetry}\label{remarks}

In treating the  vector mesons in nuclear processes, scale symmetry played an important but background role with the density dependence of the scaling factor $\phi^\ast$.  I would now like to discuss how hidden scale symmetry emerges at various different length scales in nuclear dynamics. I can do this only in effective field theory given that that's the only viable approach consistent with Weinberg's Folk Theorem on quantum effective field theory that I can resort to in nuclear matter. Nuclear physics involves various length scales from low to very high which necessitates various EFTs, with no obvious constraints between them.

$\bullet$ {\bf Unitarity limit}:
At very low energy -- and density/temperature --, say,  at a scale much less than the pion mass, one can integrate out the pion from chiral Lagrangian and be left with what's called ``pionless EFT."  It is fairly well established that the pionless EFT works for physics of light nuclei with the ``unitarity limit" in action~\cite{bira-unitarity}. (See also \cite{CS-heavynuclei} in heavier nuclei.)  What governs there is  non-relativistic conformal invariance in nuclear interactions involving only nucleons as the relevant degrees of freedom~\cite{NR-conformal}. It is an {\it emerging} symmetry from the interactions not obviously related to the scale invariance that may or may not figure in QCD. The system with such conformal symmetry may be a nucleus as in \cite{bira-unitarity} or  an ``un-stuff" where the ``un-stuff" could be an unparticle~\cite{georgi}, an unnucleus~\cite{son}, or un-matter~\cite{un-fermi} etc. 

$\bullet$ {\bf Quenched $g_A$ in nuclei}:
A case where scale invariance gets ``unhidden" in nuclear systems at a scale where pions and heavier-mesons become relevant degrees of freedom is the ``quenched $g_A$" in nuclear Gamow-Teller transitions~\cite{gA}.  Here the EFT that figures is the Landau-Migdal fixed-point theory. Unlike the case of pionless EFT where the large scattering length in NN interactions sets in for conformal invariance, here the mechanism for conformality is a lot subtler. It involves many-body correlations. This problem is addressed in the $Gn$EFT involving the $\psi\chi$HLS Lagrangian. Recall that quantum scale symmetry is implemented in the $Gn$HLS via the conformal compensator to HLS fields dual to the QCD variables. It was first shown in \cite{friman-rho} that doing the relativistic mean-field approximation with the $Gn$EFT is equivalent to doing the Landau Fermi-liquid fixed point approximation on the Fermi surface~\cite{shankar}.  With the incorporation of topology change that implements the hadron-quark continuity at $n\sim 3n_0$, this approach can go beyond the density regime where standard $\chi$EFT breaks down and has been successfully applied to compact-star physics~\cite{MR-review}.  This formalism has been found to be suitable to address the $g_A$ problem

In $\psi\chi$HLS, the axial-current coupling to the nucleon is scale-invariant. In the limit $N_c\to\infty$ and $\bar{N}\to\infty$ where $\bar{N}=k_f/(\Lambda-k_F)$, a deceptively simple result is obtained~\cite{friman-rho,gA} for the nucleon on the Fermi surface making the axial transition with zero momentum transfer. It is given by 
\be
g_A^{\rm Landau}=g_A (1-\kappa)^{-2}\label{snc}
\ee
where $g_A=1.276...$ is the axial-current coupling constant for the neutron beta decay and
\be
\kappa= \frac 13 \phi^\ast \tilde{F}_1^\pi.
\ee
Here $\tilde{F}_1^\pi$ is the pionic contribution to the Landau mass and $\phi^\ast$ is the density scaling factor (\ref{Phistar}). Putting the known numerical values, one gets
\be
g_A^{\rm Landau}\approx 1.0.   \label{landau}
\ee 
 The product  $\phi^\ast \tilde{F}_1^\pi$ turns to be more or less independent of  the density around $n_0$ due to the cancelation between the two terms. The quality of the prediction is of the same accuracy as the soft theorems like the Goldberger-Treiman relation, which is known to be good to $\sim 5\%$. One cannot make any more  precise estimate of the error bars involved. But it makes a clear qualitative physics statement. Now the key observation is that the quantity (\ref{landau}) can be equated to the ``quenched $g_A$" in the ``extreme single-particle shell model" (ESPSM) model that can be applied to the doubly closed-shell system~\cite{gA}.  There is then a quenching factor $q=1/1.276\approx 0.78$.  To give a meaning to (\ref{landau}), it should be  recalled that in the dilaton-limit fixed point, $g_A\to 1$ is what is expected as a signal for  the restoration of  scale invariance. Thus (\ref{landau}) and the DLFP $g_A$ are related. Suppose one had the ``exact" wave-functions of the initial and final states for the transition. Then no quenching of $g_A$ should be needed. This means that it's the full nuclear correlations (a.k.a. nuclear renormalizations) that would restore scale invariance. The scale invariance must therefore be hidden in the quenching factor. This is consistent with the notion of ``quantum scale invariance"\footnote{That is,   scale invariance is hidden and appears explicitly only after full renormalizations~\cite{QSI}.}.   No symmetry in nature is perfect, so there are certainly uncertainties in this prediction, a matter that could be ameliorated by future precision experiments~\cite{multifarious}. 
 
%$\bullet$ {\bf Pseudo-conformal sound speed}:  
 Finally as already mentioned, the prediction for the pseudo-conformal sound speed $v_{pcs}^2/c^2\approx 1/3$ is a manifestation of hidden conformal structure involving topology change emerging at higher densities  in compact stars~\cite{MR-review}. Together with its role at the chiral phase transition, there is the pervading scale symmetry hidden from very low to very high density.

\section{Conclusions}
There are two qualitatively reliable statements that one could make from the arguments presented in this note. 

One is that a variety of intricate connections exist to indicate that for the {\it correct} description of the properties of vector mesons near chiral restoration, it is {\it indispensable}, in agreement with \cite{kitano}, that the fractional quantum Hall pancakes be properly incorporated.  This has been done so far neither at high density nor at high temperature. This would be specially relevant to the dilepton experiments in relativistic heavy-ion collisions searching for signals of dropping vector-meson masses at high temperature. 

The other is that at various different length scales inevitably involved in nuclear processes, scale invariance is either broken or  hidden and would emerge {\it only} when full renormalizations, both fundamental (QCD) and many-body nuclear correlations, are properly taken into account. This is of course too tall an order to fully satisfy. Furthermore the emergent scale symmetries are specific to the EFTs at different scales involved and may not necessarily be linked to each other.  This means that they could manifest  differently at different scales as seen for instance in the unitarity limit in light nuclei, in the quenched $g_A$ in nuclear matter and in the pseudo-conformal sound speed in massive neutron-star matter.

 \vfil

\end{document}